\documentclass[twocolumn, 10pt, prl, aps, showpacs, reprint, amsmath, amssymb, superscriptaddress, nofootinbib]{revtex4-1}

\usepackage[]{graphicx} % Include figure files
\usepackage[]{subfigure}

\newcommand{\pbyp}[2]{\frac{\partial #1}{\partial #2}}

\newcommand{\ex}{\vec{e}_x}
\newcommand{\ey}{\vec{e}_y}

\newcommand{\sinc}{\mathrm{sinc}}

\newcommand{\alphacrit}{\alpha_{\mathrm{crit}}}

\newcommand{\alphainc}{\alpha_\mathrm{inc}}
\newcommand{\phiout}{\varphi_\mathrm{out}}

\newcommand{\Psimod}{\Psi_\mathrm{mod}}

\newcommand{\reffig}[1]{\mbox{Fig.~\ref{#1}}}

\newcommand{\refeq}[1]{\mbox{Eq.~(\ref{#1})}}

\newcommand{\reftab}[1]{\mbox{Table \ref{#1}}}
\renewcommand{\Re}[1]{\mathrm{Re}\left(#1\right)}
\renewcommand{\Im}[1]{\mathrm{Im}\left(#1\right)}

\begin{document}

\title{Origin of emission from square-shaped organic microlasers}

\author{S. Bittner}
\affiliation{Laboratoire de Photonique Quantique et Mol{\'e}culaire, {\'E}cole Normale Sup{\'e}rieure de Cachan, Centrale Sup\'elec, CNRS, Universit\'e Paris-Saclay, F-94235 Cachan, France}
\affiliation{Department of Applied Physics, Yale University, New Haven, Connecticut 06520, USA}
\author{C. Lafargue}
\affiliation{Laboratoire de Photonique Quantique et Mol{\'e}culaire, {\'E}cole Normale Sup{\'e}rieure de Cachan, Centrale Sup\'elec, CNRS, Universit\'e Paris-Saclay, F-94235 Cachan, France}
\author{I. Gozhyk}
\affiliation{Laboratoire de Photonique Quantique et Mol{\'e}culaire, {\'E}cole Normale Sup{\'e}rieure de Cachan, Centrale Sup\'elec, CNRS, Universit\'e Paris-Saclay, F-94235 Cachan, France}
\affiliation{Surface du Verre et Interfaces, UMR 125 CNRS/Saint-Gobain Recherche, 39 quai Lucien Lefranc, F-93303 Aubervilliers, France}
\author{N. Djellali}
\author{L. Milliet}
\author{D. T. Hickox-Young}
\affiliation{Laboratoire de Photonique Quantique et Mol{\'e}culaire, {\'E}cole Normale Sup{\'e}rieure de Cachan, Centrale Sup\'elec, CNRS, Universit\'e Paris-Saclay, F-94235 Cachan, France}
\author{C. Ulysse}
\affiliation{Laboratoire de Photonique et Nanostructures, CNRS UPR20, Route de Nozay, F-91460 Marcoussis, France}
\author{D. Bouche}
\affiliation{Centre de Math{\'e}matiques et de leurs Applications, {\'E}cole Normale Sup{\'e}rieure de Cachan, CNRS, Universit\'e Paris-Saclay, F-94235 Cachan, France}
\author{R. Dubertrand}
\affiliation{D{\'e}partement de Physique, Universit{\'e} de Li\`ege, 4000 Li\`ege, Belgium}
\author{E. Bogomolny}
\affiliation{Universit{\'e} Paris-Sud, Laboratoire de Physique Th{\'e}orique et Mod\`eles Statistiques, CNRS UMR 8626, Orsay, F-91405, France}
\author{J. Zyss}
\author{M. Lebental}
\email{lebental@lpqm.ens-cachan.fr}
\affiliation{Laboratoire de Photonique Quantique et Mol{\'e}culaire, {\'E}cole Normale Sup{\'e}rieure de Cachan, Centrale Sup\'elec, CNRS, Universit\'e Paris-Saclay, F-94235 Cachan, France}

\begin{abstract}
The emission from open cavities with non-integrable features remains a challenging problem of practical as well as fundamental relevance. Square-shaped dielectric microcavities provide a favorable case study with generic implications for other polygonal resonators. We report on a joint experimental and theoretical study of square-shaped organic microlasers exhibiting a far-field emission that is strongly concentrated in the directions parallel to the side walls of the cavity. A semiclassical model for the far-field distributions is developed that is in agreement with even fine features of the experimental findings. Comparison of the model calculations with the experimental data allows the precise identification of the lasing modes and their emission mechanisms, providing strong support for a physically intuitive ray-dynamical interpretation. Special attention is paid to the role of diffraction and the finite side length.
\end{abstract}


\maketitle

\section{Introduction}
Semiclassical physics emerged during the development of quantum mechanics, almost one century ago, to account for the transition from wave physics in the quantum regime to classical mechanics \cite{Brack2003}. Today, semiclassical physics plays an essential role in quantum chaos \cite{StoeckmannBuch2000}, and its methods are applied in virtually any field that features wave dynamics, including acoustics \cite{Tanner2007}, electromagnetism \cite{StoeckmannBuch2000}, hydrodynamics \cite{Kudrolli2001}, and loop quantum gravity \cite{Bianchi2011}. We demonstrate the power of these methods when applied to optical microresonators by studying square-shaped polymer-based microlasers. Their emission properties are little understood for two reasons. Firstly, the square resonator with Dirichlet boundary conditions is a standard example of a separable system, whereas the dielectric square resonator is nonseparable and hence nonintegrable due to the diffraction at the dielectric corners. This remains an open problem in mathematical physics with tremendous impact on radar or telecommunication applications \cite{Gennarelli2015}. Secondly, it was observed that these lasers emit very narrow lobes in only a few directions. Directional emission from microlasers has been intensely investigated due to possible applications \cite{Wiersig2011b} and observed for, e.g., lima\c{c}on- and stadium-shaped microlasers. The underlying mechanism is understood in terms of their chaotic ray dynamics \cite{Schwefel2004, Wiersig2008}. However, these explanations cannot be applied to polygonal resonators since they do not exhibit chaotic ray dynamics. % paragraph

While square and hexagonal microresonators and \mbox{-lasers} have been extensively studied, their far-field distributions have been scarcely investigated \cite{Wiersig2003, Yang2007c, Chen2009, Dai2009a}. This is partly because rounded corners significantly influence the resonance frequencies and field distributions \cite{Boriskina2005, Wiersig2003, Dietrich2012, Kudo2013}, and it is technologically challenging to fabricate sharp corners. Here we investigate the far-field emission of microlasers with sharp corners (i.e., with a radius of curvature much smaller than the wavelength). A semiclassical model for the dielectric square \cite{Bittner2013b, Bittner2014a} is used to predict the far-field distributions, which are in very good agreement with numerical calculations. Careful comparison of the measured far-field distributions with the model allows to identify the lasing modes and understand the mechanism for their high directionality.

\section{Experiments}

% Figure 1: SEM and camera images
\begin{figure}[tb]
\begin{center}
\includegraphics[width = 8 cm]{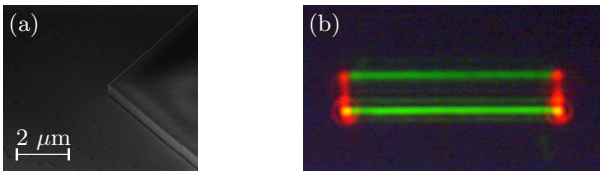}
\end{center}
\caption{(Color online) (a) Scanning electron microscope image of a square microlaser. (b) Perspective photograph in real colors of a lasing square cavity with side length $120~\mu$m. The side walls parallel to the camera axis emit red laser light, whereas the side walls perpendicular to it scatter the green pump laser.}
\label{fig:images}
\end{figure}

The microlasers consisted of a PMMA\footnote{Poly(methyl methacrylate) PMMA A6 resist by Microchem.} polymer matrix doped by $5$~wt\% DCM\footnote{4-(Dicyanomethylene)-2-methyl-6-(4-dimethylaminostyryl)-4Hpyran by Exciton.} laser dye that was deposited in a $650$~nm-thick layer on a Si/SiO$_2$ ($2~\mu$m) wafer by spin-coating. Square cavities with side length ranging from $a = 80$ to $200~\mu$m were engraved by electron-beam lithography \cite{Lozenko2012}, which makes it possible to obtain two-dimensional cavities with nanoscale precision [see \reffig{fig:images}(a)]. % paragraph

The microlasers were pumped by a pulsed, frequency-doubled Nd:YAG laser ($532$~nm, $0.5$~ns, $10$~Hz). The short pulses and low duty cycle help to avoid problems from heating and quenching due to dark states. The pump beam impinged vertically and it covered the whole cavity uniformly The lasing emission was collected in the sample plane by a lens and transferred to a spectrometer by an optical fiber. It was polarized parallel to the sample plane. The far-field intensity distributions were measured as a function of the azimuthal angle $\varphi$ by rotating the cavity.

% Figure 2: Spectrum of a square microlaser (a=120 µm) and its FT
\begin{figure}[tb]
\begin{center}
\includegraphics[width = 8 cm]{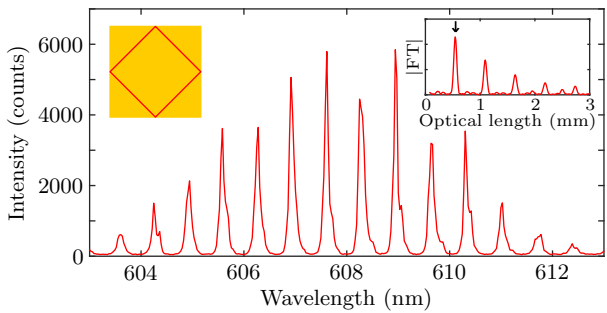}
\end{center}
\caption{Experimental spectrum of a square microlaser with $120~\mu$m side length at $\varphi = 0^\circ$ at about $2.5$ times the threshold intensity ($I_\mathrm{thres} = 2.5$~MW cm$^{-2}$) and with linearly polarized pump beam. The left inset shows the diamond periodic orbit in the square billiard. The right inset shows its Fourier transform. The arrow indicates the optical length of the diamond orbit.}
\label{fig:specExp}
\end{figure}

A typical spectrum is shown in \reffig{fig:specExp}. It features a sequence of equidistant peaks. Its Fourier transform (right inset of \reffig{fig:specExp}) shows that the free spectral range (FSR) of the spectrum corresponds to the optical length of the diamond orbit \cite{Lebental2007, Bogomolny2011}. Therefore it is expected that the observed lasing modes are localized on trajectories with an angle of incidence close to $45^\circ$. % paragraph

% Figure 3: Experimental far-field intensity distribution
\begin{figure}[tb]
\begin{center}
\includegraphics[width = 8 cm]{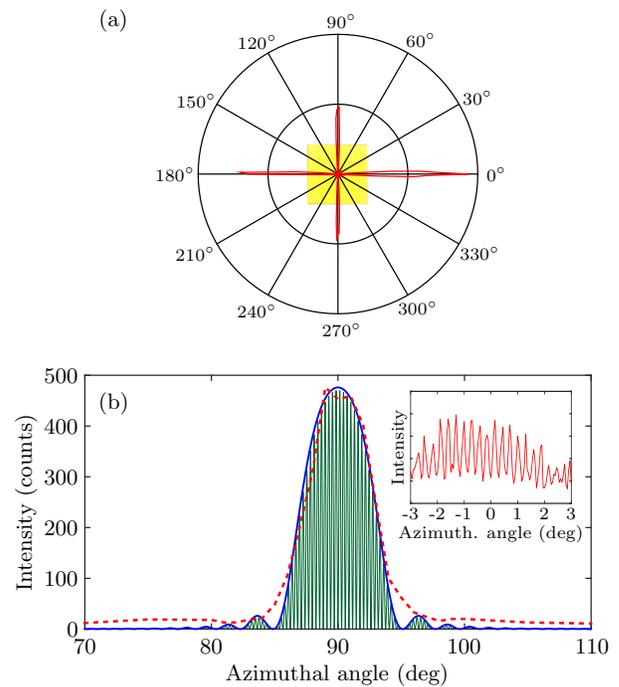}
\end{center}
\caption{(a) Measured far-field intensity distribution of a single resonance at $\lambda = 606.7$~nm for circularly polarized pump beam. The photograph of the cavity indicates its orientation. (b) Magnification around the lobe at $90^\circ$. The dashed red line is the measured intensity distribution, the solid blue line the fitted intensity envelope, and the thin green line the corresponding calculated intensity distribution. The inset shows a measurement with higher angular resolution and linearly polarized pump around $0^\circ$.}
\label{fig:FFexp}
\end{figure}

A typical far-field intensity distribution of a single resonance is presented in \reffig{fig:FFexp}(a). It exhibits four emission lobes in the directions parallel to the side walls. The emission lobes are very narrow with a full width at half maximum of about $6^\circ$ [see \reffig{fig:FFexp}(b)]. Figure \ref{fig:images}(b) shows an image of the lasing cavity taken by a camera with a high-magnification zoom lens. It evidences that the lasing emission (red) is emitted from the side walls parallel to the emission direction, whereas the side walls perpendicular to it do not emit but only scatter the green pump laser. Since the emission is at a grazing angle, it must stem from rays impinging on the cavity side walls with an angle of incidence very close to the critical one, $\alphacrit = \arcsin(1 / n) = 41.8^\circ$, where $n = 1.5$ is the refractive index \cite{Lebental2007}. This is in agreement with the previous observation that the ray trajectories on which the lasing modes are based have an angle of incidence close to $45^\circ$. % paragraph

\section{Model calculations}

% Figure 4: Diagram of rays in the resonator and outside
\begin{figure}[tb]
\begin{center}
\includegraphics[height = 4.5 cm]{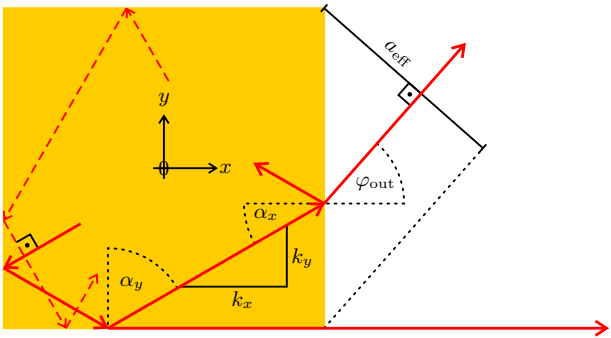}
\end{center}
\caption{Sketch of rays with momentum vectors $\pm k_x \ex \pm k_y \ey$ (solid red lines). Their angles of incidence on the vertical (horizontal) side walls are $\alpha_x$ ($\alpha_y = \pi/2 - \alpha_x$). The angle of refraction is $\phiout$. The dashed red lines indicate rays with momentum vectors rotated by $90^\circ$, $\pm k_y \ex \pm k_x \ey$}
\label{fig:rayDiagram}
\end{figure} 

The far-field distributions are calculated using the semiclassical model introduced in Ref.~\cite{Bittner2013b, Bittner2014a} to identify the observed lasing modes. The model is based on the observation that the wave functions are localized on classical tori and are thus composed of eight plane waves $\exp\{i (k_x x +  k_y y) \}$. Their directions are related to each other by the symmetry operations of the $C_{4v}$ point group. The corresponding ray trajectories in \reffig{fig:rayDiagram} are given by the momentum vectors $\pm k_x\vec e_x \pm k_y\vec e_y$ and their rotations by 90$^{\circ}$, $\pm k_y\vec e_x \pm k_x\vec e_y$. % paragraph

To quantize the spectrum, the simplest approach is to separate the system along the $x$ and $y$ directions, and write two phase loop conditions analogous to that of a Fabry-P\'erot cavity \cite{Bittner2013b, Bittner2014a},
\begin{equation}
\label{eq:resCond}
\begin{array}{lcl}
r^2(\alpha_x) \, e^{ik_x 2 a} & = & 1 \, , \\
r^2(\alpha_y) \, e^{ik_y2a} & = & 1 \, .
\end{array}
\end{equation}
The main difference from a Fabry-P\'erot cavity are the Fresnel coefficients $r$ corresponding to an incidence with angles $\alpha_{x, y} = \arctan[\Re{k_{y, x}}/\Re{k_{x, y}}]$ instead of $0^\circ$. The momentum vector components $k_{x,y}$ can thus be formally written as
\begin{equation}
\begin{array}{rcl}
k_x & = & \{\pi m_x + i \ln[r(\alpha_x)] \} / a \, , \\
k_y & = & \{\pi m_y + i \ln[r(\alpha_y)] \} / a \, .
\end{array}
\end{equation}
The resonance wave number is $k = 2 \pi / \lambda = (k_x^2 + k_y^2)^{1/2} / n$ where $\lambda$ is the free-space wavelength. Note that $k_{x, y}$ are in general complex-valued. All wave functions $\Psi(x, y)$ are obtained by adding up the $8$ plane waves with their correct momentum vectors and signs. For transverse magnetic (TM) [transverse electric (TE)] polarization, the electric (magnetic) field is parallel to the plane of the cavity, and $\Psi$ corresponds to the $z$ component of the electric (magnetic) field. The resonances [i.e., solutions of \refeq{eq:resCond}] are labeled by their quantum numbers $m_{x, y}$ and their symmetries $s_1$ ($s_2$) with respect to the diagonal $x = y$ ($x = -y$) as $(m_x, m_y, s_1 s_2)$, where $s_{1, 2} = +1$ ($s_{1, 2} = -1$) means a symmetric (antisymmetric) wave function. There are six symmetry classes corresponding to the different combinations of $s_{1, 2}$ and the quantum numbers that are labeled by the Mulliken symbols $A_{1, 2}$, $B_{1, 2}$ and $E$ (see \reftab{tab:dlmWFs}). % paragraph

% Table 01: Quantum numbers, symmetry classes and wave functions for the DLM
\begin{table*}[tb]
\caption{Symmetry classes, quantum numbers and model WFs (adapted from Ref.~\cite{Bittner2014a}).}
\label{tab:dlmWFs}
\vspace{3 mm}
\begin{center}
\begin{tabular}{c|c|c|c|c|l}
\hline
\hline
Diagonal & Horizontal/vertical & Parity of & Parity of & Mulliken & Model wave function \\
symmetry & symmetry & $m_x + m_y$ & $m_x \cdot m_y$ & symbol & \\
\hline
$(++)$ & $+$ & Even & Even & $A_1$ & $\Psimod(x, y) = \Psi_0 [ \cos(k_x x) \cos(k_y y) + \cos(k_y x) \cos(k_x y) ]$ \\
$(--)$ & $+$ & Even & Even & $B_2$ & $\Psimod(x, y) = \Psi_0 [ \cos(k_x x) \cos(k_y y) - \cos(k_y x) \cos(k_x y) ]$ \\
$(++)$ & $-$ & Even & Odd & $B_1$ & $\Psimod(x, y) = \Psi_0 [ \sin(k_x x) \sin(k_y y) + \sin(k_y x) \sin(k_x y) ]$ \\
$(--)$ & $-$ & Even & Odd & $A_2$ & $\Psimod(x, y) = \Psi_0 [ \sin(k_x x) \sin(k_y y) - \sin(k_y x) \sin(k_x y) ]$ \\
$(+-)$ & None & Odd & Even & $E$ & $\Psimod(x, y) = \Psi_0 [ \sin(k_x x) \cos(k_y y) + \cos(k_y x) \sin(k_x y) ]$ \\
$(-+)$ & None & Odd & Even & $E$ & $\Psimod(x, y) = \Psi_0 [ \sin(k_x x) \cos(k_y y) - \cos(k_y x) \sin(k_x y) ]$ \\
\hline
\hline
\end{tabular}
\end{center}
\end{table*}

Green's identity is used to infer the far-field distribution \cite{Yang2007c}. The derivation is detailed in the Supplementary Materials along with the comparison to numerical simulations (see Figs.~\ref{fig:specNum} and \ref{fig:FFnum}). The full expression for the far-field distribution of an $A_2$ mode is given in \refeq{eq:ffA2}. It consists of $8$ terms like
\begin{equation} \label{eq:sincTerm}
\begin{array}{c}
\sinc\{(k_y - k \sin{\varphi}) \frac{a}{2} \} [\mu k_x \frac{a}{2} \cos(k_x \frac{a}{2}) \sin(k \frac{a}{2} \cos{\varphi}) \\ - k \frac{a}{2} \cos{\varphi} \cos(k \frac{a}{2} \cos{\varphi}) \sin(k_x \frac{a}{2})]
\end{array}
\end{equation}
where $\mu = 1$ ($\mu = 1 / n^2$) for TM (TE) polarization and $\sinc(x) = \sin(x) / x$. Each such term corresponds to emission in the two directions $\phiout$ defined by the roots of the argument of the $\sinc$ function. If the angle of incidence of the corresponding rays inside the resonator is smaller than $\alphacrit$, then $\phiout$ is simply given by Snell's law (see \reffig{fig:rayDiagram}), and the emission lobe is called \emph{refractive}. Then the $\sinc$ term is in fact the diffraction pattern of a plane wave going through a slit with a width $a_\mathrm{eff}$ that is the projection of the side of the square on the emission direction (see \reffig{fig:rayDiagram}). When $\phiout$ is complex, corresponding to rays with angle of incidence larger than $\alphacrit$, the lobe is called \emph{nonrefractive} and is emitted parallel to the side wall. The nonrefractive lobes are jagged and much broader. They have a much smaller amplitude than the refractive lobes due to the finite imaginary part of $\phiout$. While no energy is transmitted when a plane wave impinges on an infinite dielectric interface with an angle larger than the critical one (i.e., it is totally reflected), this is not the case for an interface of finite length as illustrated by the existence of the nonrefractive emission lobes \cite{Wiersig2003}, which leads to a modification of the reflection and transmission coefficients. % paragraph

Two different cases of far-field diagrams can appear for $n = 1.5 > \sqrt{2}$, depending on $\alphainc = \min\{\alpha_x, \alpha_y\}$. If $\alphainc < \alphacrit$, the plane waves impinging on a side wall with angle $\alphainc$ escape refractively, whereas those impinging with $\pi / 2 - \alphainc$ are totally reflected. Hence $8$ refractive and $8$ nonrefractive lobes are observed, where the latter are typically negligible compared to the former. If $\alphainc \geq \alphacrit$, all plane waves are totally reflected, and $16$ nonrefractive lobes are observed, though their directions are four-fold degenerate.

\section{Comparison with experiment}

% Figure 5: Model spectrum lambda-vs-alpha
\begin{figure}[tb]
\begin{center}
\includegraphics[width = 8 cm]{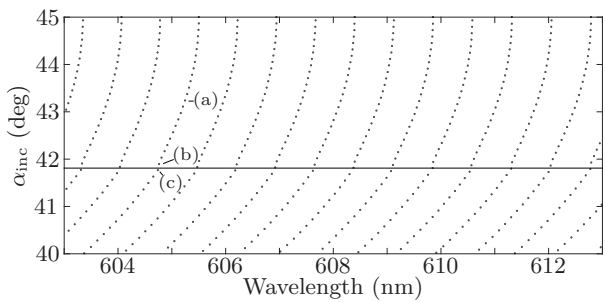}
\end{center}
\caption{Model calculation of the TE modes of a $a = 120~\mu$m square with $n = 1.5$. The angle of incidence is plotted with respect to the resonance wavelength. The horizontal line indicates the critical angle. The model far-field intensity distributions of the modes marked (a), (b), and (c) are shown in \reffig{fig:FFmod}.}
\label{fig:specMod}
\end{figure}

We now compare the model with experiments. A spectrum generated by the model and corresponding to the experimental conditions is shown in \reffig{fig:specMod}. The resonances are arranged in branches. Each of them consists of modes with identical longitudinal quantum number $m = m_x + m_y$ and increasing transverse quantum number $p = |m_x - m_y| / 2$ as $\alphainc$ decreases \cite{Bittner2014a}. The horizontal distance between these branches agrees well with the experimentally observed FSR, corresponding to the length of the diamond orbit. Therefore we conclude that the observed lasing resonances belong to different branches, i.e., each has a different $m$.

% Figure 6: Examples of model far-field intensity distributions for realistic parameters
\begin{figure}[tb]
\begin{center}
\includegraphics[width = 8 cm]{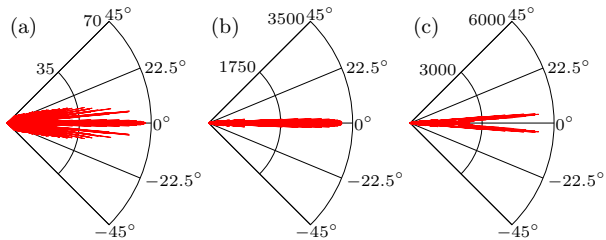}
\end{center}
\caption{Model calculations of the far-field intensity distributions for a $a = 120~\mu$m square with $n = 1.5$. The modes are (a) TE$(433, 407, --)$, (b) TE$(443, 397, --)$, and (c) TE$(444, 396, --)$.}
\label{fig:FFmod}
\end{figure}

We use the far-field intensity distributions to infer $\alphainc$ and thus their transverse quantum numbers. The model far-field patterns of three representative modes are presented in \reffig{fig:FFmod}. They are labeled by (a), (b), and (c) in \reffig{fig:specMod}. Mode (c) is located just below the critical angle ($p = 24$) and features two refractive emission lobes between $\varphi = -45^{\circ}$ and $45^{\circ}$, while its nonrefractive lobes are not visible due to their negligible amplitude. Since the measured far-field patterns feature only $4$ lobes in the total range of $360^\circ$, such modes with $\alphainc < \alphacrit$ can be excluded. Mode (b) with $p = 23$ is located just above the critical angle. The refractive lobes have become nonrefractive and merged into four narrow lobes parallel to the sides. This is precisely the kind of far-field pattern observed experimentally. Mode (a) finally is well above the critical angle ($p = 13$). The nonrefractive lobes have become very broad and jagged unlike what was observed experimentally. Also their amplitude has decreased significantly. This development continues with increasing $\alphainc$. The qualitative comparison with the measured far-field intensity distribution shows that the experimental lasing modes are only consistent with mode (b), i.e., an angle of incidence just above the critical angle. This is confirmed by the fits described in the following. % paragraph

The experimental far-field distribution measured with an angular resolution of 1$^{\circ}$ as in \reffig{fig:FFexp} is compared to the envelope of the model far-field distributions. This is detailed in the Supplementary Materials. The only two independent parameters are the amplitude and $\alphainc$, which is considered a continuous variable due to the high resonance density. The fit of the lobe yields $\alphainc^\mathrm{(fit)} = 41.87^\circ$ and is plotted in \reffig{fig:FFexp}(b). The agreement is excellent. Similar results are obtained for the other lobes and the other resonances as well as squares with various sizes, yielding values of $\alphainc^\mathrm{(fit)}$ from $41.81^\circ$ to $41.96^\circ$. As expected from the qualitative considerations in the previous paragraph and the photograph in \reffig{fig:images}(b), the angle of incidence is slightly above the critical angle. % paragraph

In fact, the model predicts far-field distributions with a very quickly oscillating substructure that stems from the terms in the square brackets in \refeq{eq:sincTerm}, namely the $\sin$ and $\cos$ functions with the argument $\phi = k a \cos(\varphi) / 2$. The actual model far-field intensity distribution for the fitted parameters is shown as thin green line in \reffig{fig:FFexp}(b). To confirm this prediction, the angular resolution of the setup was improved by putting a slit in front of the collection lens. The far-field intensity distribution measured this way is shown in the inset of \reffig{fig:FFexp}(b). It features oscillations with a period of $(0.29 \pm 0.01)^\circ$, while the expected period is $0.29^\circ$ according to \refeq{eq:sincTerm}. The agreement is excellent for different wavelengths and cavity sizes. % paragraph

Systematic measurements with increased angular resolution will allow for a better determination of the symmetry of the lasing resonances. Furthermore, the symmetry of the far-field distributions can be changed by the polarization of the pump laser due to the fluorescence anisotropy of the laser dye \cite{Gozhyk2012}: when the square microlaser is pumped with linear polarization parallel to the $x$ axis [instead of circular polarization as in \reffig{fig:FFexp}(a)], the emission is almost bidirectional with strong emission lobes at $\varphi = 0^\circ$ and $180^\circ$, whereas the emission lobes at $90^\circ$ and $270^\circ$ are strongly suppressed (not shown). This effect is a consequence of the lower degree of symmetry induced by the pump polarization. It can be explained by the formation of coherent superpositions of degenerate mode pairs like the $A_1$ and $B_2$ modes. It should be noted that the lasing modes and hence far-field emission of microlasers can also be controlled by localized or selective pumping \cite{Ben-Messaoud2005, Bachelard2012, Liew2014}, whereas controlling it only with the pump polarization is an experimentally simpler approach. These topics are, however, beyond the scope of this article and will be explored in a future publication.

\section{Conclusions}
The far-field intensity distributions of square-shaped organic microlasers were investigated and revealed highly directional emission in the four directions parallel to the cavity sides. Sharp corners were essential for obtaining this high directionality because rounded corners significantly change the emission behavior \cite{Wiersig2003}. Analytical formulas for the far-field distributions based on a semiclassical model \cite{Bittner2013b, Bittner2014a} were developed and showed excellent agreement both with experimental data and with numerical simulations. They could be easily adapted to other polygonal cavities like hexagon or rectangles \cite{Yang2007c} and other refractive indices. The comparison between the model and experiments evidences that the lasing modes are based on trajectories with an angle of incidence almost equal to the critical angle that hence leave the resonator at a grazing angle. Interestingly, these modes do not feature the highest quality factors; these would be modes with $\alphainc$ larger than $\alphacrit$. Maybe these modes actually lase, but are not observed because of their lower amplitude in the far-field compared to other modes. Or they may actually have higher losses due to diffraction at the substrate interface \cite{Lozenko2012}. % paragraph

The semiclassical model does not take into account diffraction by the corners, whereas it is precisely this effect which renders the system nonintegrable. While the model calculations agree very well with the observed spectrum and far-field distributions, there are some minor observations that cannot be explained by them. For example, some light is emitted into other directions than the four main emission lobes due to diffraction at the corners, but it is much less intense\footnote{For a video of the rotating square microlaser please see the Supplementary Material at http://dx.doi.org/10.1209/0295-5075/113/54002}. This means that diffraction by the corner plays a secondary, but not negligible role for the emission from polygonal microlasers. Studying these effects would allow more insight into the open physical-mathematical problem of diffraction by a dielectric wedge.

\begin{acknowledgments}
The numerical calculations are based on a code developed by C.~Schmit. S.~B.\ gratefully acknowledges funding from the European Union Seventh Framework Programme (FP7/2007-2013) under Grant No.\ 246.556.10. This work was supported by a public grant from the Laboratoire d'Excellence Physics Atom Light Matter (LabEx PALM) overseen by the French National Research Agency (ANR) as part of the Investissements d'Avenir program (Reference No.\ ANR-10-LABX-0039).
\end{acknowledgments}

% The bibiliography
% -----------------
%\bibliography{Refs}

\begin{thebibliography}{27}%
\makeatletter
\providecommand \@ifxundefined [1]{%
 \@ifx{#1\undefined}
}%
\providecommand \@ifnum [1]{%
 \ifnum #1\expandafter \@firstoftwo
 \else \expandafter \@secondoftwo
 \fi
}%
\providecommand \@ifx [1]{%
 \ifx #1\expandafter \@firstoftwo
 \else \expandafter \@secondoftwo
 \fi
}%
\providecommand \natexlab [1]{#1}%
\providecommand \enquote  [1]{``#1''}%
\providecommand \bibnamefont  [1]{#1}%
\providecommand \bibfnamefont [1]{#1}%
\providecommand \citenamefont [1]{#1}%
\providecommand \href@noop [0]{\@secondoftwo}%
\providecommand \href [0]{\begingroup \@sanitize@url \@href}%
\providecommand \@href[1]{\@@startlink{#1}\@@href}%
\providecommand \@@href[1]{\endgroup#1\@@endlink}%
\providecommand \@sanitize@url [0]{\catcode `\\12\catcode `\$12\catcode
  `\&12\catcode `\#12\catcode `\^12\catcode `\_12\catcode `\%12\relax}%
\providecommand \@@startlink[1]{}%
\providecommand \@@endlink[0]{}%
\providecommand \url  [0]{\begingroup\@sanitize@url \@url }%
\providecommand \@url [1]{\endgroup\@href {#1}{\urlprefix }}%
\providecommand \urlprefix  [0]{URL }%
\providecommand \Eprint [0]{\href }%
\providecommand \doibase [0]{http://dx.doi.org/}%
\providecommand \selectlanguage [0]{\@gobble}%
\providecommand \bibinfo  [0]{\@secondoftwo}%
\providecommand \bibfield  [0]{\@secondoftwo}%
\providecommand \translation [1]{[#1]}%
\providecommand \BibitemOpen [0]{}%
\providecommand \bibitemStop [0]{}%
\providecommand \bibitemNoStop [0]{.\EOS\space}%
\providecommand \EOS [0]{\spacefactor3000\relax}%
\providecommand \BibitemShut  [1]{\csname bibitem#1\endcsname}%
\let\auto@bib@innerbib\@empty
%</preamble>
\bibitem [{\citenamefont {Brack}\ and\ \citenamefont
  {Bhaduri}(2003)}]{Brack2003}%
  \BibitemOpen
  \bibfield  {author} {\bibinfo {author} {\bibfnamefont {M.}~\bibnamefont
  {Brack}}\ and\ \bibinfo {author} {\bibfnamefont {R.~K.}\ \bibnamefont
  {Bhaduri}},\ }\href@noop {} {\emph {\bibinfo {title} {Semiclassical
  Physics}}}\ (\bibinfo  {publisher} {Westview Press},\ \bibinfo {address}
  {Oxford},\ \bibinfo {year} {2003})\BibitemShut {NoStop}%
\bibitem [{\citenamefont {St{\"o}ckmann}(2000)}]{StoeckmannBuch2000}%
  \BibitemOpen
  \bibfield  {author} {\bibinfo {author} {\bibfnamefont {H.-J.}\ \bibnamefont
  {St{\"o}ckmann}},\ }\href@noop {} {\emph {\bibinfo {title}
  {Quantum~Chaos:~An~Introduction}}}\ (\bibinfo  {publisher} {Cam\-bridge
  University Press},\ \bibinfo {address} {Cambridge, UK},\ \bibinfo {year}
  {2000})\BibitemShut {NoStop}%
\bibitem [{\citenamefont {Tanner}\ and\ \citenamefont
  {S{\o}ndergaard}(2007)}]{Tanner2007}%
  \BibitemOpen
  \bibfield  {author} {\bibinfo {author} {\bibfnamefont {G.}~\bibnamefont
  {Tanner}}\ and\ \bibinfo {author} {\bibfnamefont {N.}~\bibnamefont
  {S{\o}ndergaard}},\ }\href {\doibase doi:10.1088/1751-8113/40/50/R01}
  {\bibfield  {journal} {\bibinfo  {journal} {J. Phys. A}\ }\textbf {\bibinfo
  {volume} {40}},\ \bibinfo {pages} {R443} (\bibinfo {year}
  {2007})}\BibitemShut {NoStop}%
\bibitem [{\citenamefont {Kudrolli}\ \emph {et~al.}(2001)\citenamefont
  {Kudrolli}, \citenamefont {Abraham},\ and\ \citenamefont
  {Gollub}}]{Kudrolli2001}%
  \BibitemOpen
  \bibfield  {author} {\bibinfo {author} {\bibfnamefont {A.}~\bibnamefont
  {Kudrolli}}, \bibinfo {author} {\bibfnamefont {M.~C.}\ \bibnamefont
  {Abraham}}, \ and\ \bibinfo {author} {\bibfnamefont {J.~P.}\ \bibnamefont
  {Gollub}},\ }\href {\doibase 10.1103/PhysRevE.63.026208} {\bibfield
  {journal} {\bibinfo  {journal} {Phys. Rev. E}\ }\textbf {\bibinfo {volume}
  {63}},\ \bibinfo {pages} {026208} (\bibinfo {year} {2001})}\BibitemShut
  {NoStop}%
\bibitem [{\citenamefont {Bianchi}\ and\ \citenamefont
  {Haggard}(2011)}]{Bianchi2011}%
  \BibitemOpen
  \bibfield  {author} {\bibinfo {author} {\bibfnamefont {E.}~\bibnamefont
  {Bianchi}}\ and\ \bibinfo {author} {\bibfnamefont {H.~M.}\ \bibnamefont
  {Haggard}},\ }\href {\doibase 10.1103/PhysRevLett.107.011301} {\bibfield
  {journal} {\bibinfo  {journal} {Phys. Rev. Lett.}\ }\textbf {\bibinfo
  {volume} {107}},\ \bibinfo {pages} {011301} (\bibinfo {year}
  {2011})}\BibitemShut {NoStop}%
\bibitem [{\citenamefont {Gennarelli}\ \emph {et~al.}(2015)\citenamefont
  {Gennarelli}, \citenamefont {Frongillo},\ and\ \citenamefont
  {Riccio}}]{Gennarelli2015}%
  \BibitemOpen
  \bibfield  {author} {\bibinfo {author} {\bibfnamefont {G.}~\bibnamefont
  {Gennarelli}}, \bibinfo {author} {\bibfnamefont {M.}~\bibnamefont
  {Frongillo}}, \ and\ \bibinfo {author} {\bibfnamefont {G.}~\bibnamefont
  {Riccio}},\ }\href {\doibase 10.1109/TAP.2014.2364305} {\bibfield  {journal}
  {\bibinfo  {journal} {IEEE Trans. Antennas Propag.}\ }\textbf {\bibinfo
  {volume} {63}},\ \bibinfo {pages} {374} (\bibinfo {year} {2015})}\BibitemShut
  {NoStop}%
\bibitem [{\citenamefont {Wiersig}\ \emph {et~al.}(2011)\citenamefont
  {Wiersig}, \citenamefont {Unterhinninghofen}, \citenamefont {Song},
  \citenamefont {Cao}, \citenamefont {Hentschel},\ and\ \citenamefont
  {Shinohara}}]{Wiersig2011b}%
  \BibitemOpen
  \bibfield  {author} {\bibinfo {author} {\bibfnamefont {J.}~\bibnamefont
  {Wiersig}}, \bibinfo {author} {\bibfnamefont {J.}~\bibnamefont
  {Unterhinninghofen}}, \bibinfo {author} {\bibfnamefont {Q.}~\bibnamefont
  {Song}}, \bibinfo {author} {\bibfnamefont {H.}~\bibnamefont {Cao}}, \bibinfo
  {author} {\bibfnamefont {M.}~\bibnamefont {Hentschel}}, \ and\ \bibinfo
  {author} {\bibfnamefont {S.}~\bibnamefont {Shinohara}},\ }\enquote {\bibinfo
  {title} {Trends in nano- and micro-cavities},}\ \ (\bibinfo  {publisher}
  {Bentham Science Publishers},\ \bibinfo {address} {Sharjah, United Arab
  Emirates},\ \bibinfo {year} {2011})\ Chap.~\bibinfo {chapter} {4}, pp.\
  \bibinfo {pages} {109--152}\BibitemShut {NoStop}%
\bibitem [{\citenamefont {Schwefel}\ \emph {et~al.}(2004)\citenamefont
  {Schwefel}, \citenamefont {Rex}, \citenamefont {Tureci}, \citenamefont
  {Chang}, \citenamefont {Stone}, \citenamefont {Ben-Messaoud},\ and\
  \citenamefont {Zyss}}]{Schwefel2004}%
  \BibitemOpen
  \bibfield  {author} {\bibinfo {author} {\bibfnamefont {H.~G.~L.}\
  \bibnamefont {Schwefel}}, \bibinfo {author} {\bibfnamefont {N.~B.}\
  \bibnamefont {Rex}}, \bibinfo {author} {\bibfnamefont {H.~E.}\ \bibnamefont
  {Tureci}}, \bibinfo {author} {\bibfnamefont {R.~K.}\ \bibnamefont {Chang}},
  \bibinfo {author} {\bibfnamefont {A.~D.}\ \bibnamefont {Stone}}, \bibinfo
  {author} {\bibfnamefont {T.}~\bibnamefont {Ben-Messaoud}}, \ and\ \bibinfo
  {author} {\bibfnamefont {J.}~\bibnamefont {Zyss}},\ }\href {\doibase
  10.1364/JOSAB.21.000923} {\bibfield  {journal} {\bibinfo  {journal} {J. Opt.
  Soc. Am. B}\ }\textbf {\bibinfo {volume} {21}},\ \bibinfo {pages} {923}
  (\bibinfo {year} {2004})}\BibitemShut {NoStop}%
\bibitem [{\citenamefont {Wiersig}\ and\ \citenamefont
  {Hentschel}(2008)}]{Wiersig2008}%
  \BibitemOpen
  \bibfield  {author} {\bibinfo {author} {\bibfnamefont {J.}~\bibnamefont
  {Wiersig}}\ and\ \bibinfo {author} {\bibfnamefont {M.}~\bibnamefont
  {Hentschel}},\ }\href {\doibase 10.1103/PhysRevLett.100.033901} {\bibfield
  {journal} {\bibinfo  {journal} {Phys. Rev. Lett.}\ }\textbf {\bibinfo
  {volume} {100}},\ \bibinfo {pages} {033901} (\bibinfo {year}
  {2008})}\BibitemShut {NoStop}%
\bibitem [{\citenamefont {Wiersig}(2003{\natexlab{a}})}]{Wiersig2003}%
  \BibitemOpen
  \bibfield  {author} {\bibinfo {author} {\bibfnamefont {J.}~\bibnamefont
  {Wiersig}},\ }\href {\doibase 10.1103/PhysRevA.67.023807} {\bibfield
  {journal} {\bibinfo  {journal} {Phys. Rev. A}\ }\textbf {\bibinfo {volume}
  {67}},\ \bibinfo {pages} {023807} (\bibinfo {year}
  {2003}{\natexlab{a}})}\BibitemShut {NoStop}%
\bibitem [{\citenamefont {Yang}\ and\ \citenamefont {Huang}(2007)}]{Yang2007c}%
  \BibitemOpen
  \bibfield  {author} {\bibinfo {author} {\bibfnamefont {Y.-D.}\ \bibnamefont
  {Yang}}\ and\ \bibinfo {author} {\bibfnamefont {Y.-Z.}\ \bibnamefont
  {Huang}},\ }\href {\doibase 10.1109/JQE.2007.897879} {\bibfield  {journal}
  {\bibinfo  {journal} {IEEE J. Quantum Electron.}\ }\textbf {\bibinfo {volume}
  {43}},\ \bibinfo {pages} {497} (\bibinfo {year} {2007})}\BibitemShut
  {NoStop}%
\bibitem [{\citenamefont {Chen}\ \emph {et~al.}(2009)\citenamefont {Chen},
  \citenamefont {Yu}, \citenamefont {Huang}, \citenamefont {Chen},
  \citenamefont {Chen},\ and\ \citenamefont {Huang}}]{Chen2009}%
  \BibitemOpen
  \bibfield  {author} {\bibinfo {author} {\bibfnamefont {R.~C.~C.}\
  \bibnamefont {Chen}}, \bibinfo {author} {\bibfnamefont {Y.~T.}\ \bibnamefont
  {Yu}}, \bibinfo {author} {\bibfnamefont {Y.~J.}\ \bibnamefont {Huang}},
  \bibinfo {author} {\bibfnamefont {C.~C.}\ \bibnamefont {Chen}}, \bibinfo
  {author} {\bibfnamefont {Y.~F.}\ \bibnamefont {Chen}}, \ and\ \bibinfo
  {author} {\bibfnamefont {K.~F.}\ \bibnamefont {Huang}},\ }\href {\doibase
  10.1364/OL.34.001810} {\bibfield  {journal} {\bibinfo  {journal} {Opt.
  Lett.}\ }\textbf {\bibinfo {volume} {34}},\ \bibinfo {pages} {1810} (\bibinfo
  {year} {2009})}\BibitemShut {NoStop}%
\bibitem [{\citenamefont {Dai}\ \emph {et~al.}(2009)\citenamefont {Dai},
  \citenamefont {Xu}, \citenamefont {Zheng}, \citenamefont {Lv},\ and\
  \citenamefont {Cui}}]{Dai2009a}%
  \BibitemOpen
  \bibfield  {author} {\bibinfo {author} {\bibfnamefont {J.}~\bibnamefont
  {Dai}}, \bibinfo {author} {\bibfnamefont {C.~X.}\ \bibnamefont {Xu}},
  \bibinfo {author} {\bibfnamefont {K.}~\bibnamefont {Zheng}}, \bibinfo
  {author} {\bibfnamefont {C.~G.}\ \bibnamefont {Lv}}, \ and\ \bibinfo {author}
  {\bibfnamefont {Y.~P.}\ \bibnamefont {Cui}},\ }\href {\doibase
  10.1063/1.3276069} {\bibfield  {journal} {\bibinfo  {journal} {Appl. Phys.
  Lett.}\ }\textbf {\bibinfo {volume} {95}},\ \bibinfo {pages} {241110}
  (\bibinfo {year} {2009})}\BibitemShut {NoStop}%
\bibitem [{\citenamefont {Boriskina}(2005)}]{Boriskina2005}%
  \BibitemOpen
  \bibfield  {author} {\bibinfo {author} {\bibfnamefont {S.~V.}\ \bibnamefont
  {Boriskina}},\ }\href {\doibase 10.1109/JQE.2005.846696} {\bibfield
  {journal} {\bibinfo  {journal} {IEEE J. Quantum Electron.}\ }\textbf
  {\bibinfo {volume} {41}},\ \bibinfo {pages} {857} (\bibinfo {year}
  {2005})}\BibitemShut {NoStop}%
\bibitem [{\citenamefont {Dietrich}\ \emph {et~al.}(2012)\citenamefont
  {Dietrich}, \citenamefont {Lange}, \citenamefont {B\"ontgen},\ and\
  \citenamefont {Grundmann}}]{Dietrich2012}%
  \BibitemOpen
  \bibfield  {author} {\bibinfo {author} {\bibfnamefont {C.~P.}\ \bibnamefont
  {Dietrich}}, \bibinfo {author} {\bibfnamefont {M.}~\bibnamefont {Lange}},
  \bibinfo {author} {\bibfnamefont {T.}~\bibnamefont {B\"ontgen}}, \ and\
  \bibinfo {author} {\bibfnamefont {M.}~\bibnamefont {Grundmann}},\ }\href
  {\doibase 10.1063/1.4757572} {\bibfield  {journal} {\bibinfo  {journal}
  {Appl. Phys. Lett.}\ }\textbf {\bibinfo {volume} {101}},\ \bibinfo {pages}
  {141116} (\bibinfo {year} {2012})}\BibitemShut {NoStop}%
\bibitem [{\citenamefont {Kudo}\ \emph {et~al.}(2013)\citenamefont {Kudo},
  \citenamefont {Suzuki},\ and\ \citenamefont {Tanabe}}]{Kudo2013}%
  \BibitemOpen
  \bibfield  {author} {\bibinfo {author} {\bibfnamefont {H.}~\bibnamefont
  {Kudo}}, \bibinfo {author} {\bibfnamefont {R.}~\bibnamefont {Suzuki}}, \ and\
  \bibinfo {author} {\bibfnamefont {T.}~\bibnamefont {Tanabe}},\ }\href
  {\doibase 10.1103/PhysRevA.88.023807} {\bibfield  {journal} {\bibinfo
  {journal} {Phys. Rev. A}\ }\textbf {\bibinfo {volume} {88}},\ \bibinfo
  {pages} {023807} (\bibinfo {year} {2013})}\BibitemShut {NoStop}%
\bibitem [{\citenamefont {Bittner}\ \emph {et~al.}(2013)\citenamefont
  {Bittner}, \citenamefont {Bogomolny}, \citenamefont {Dietz}, \citenamefont
  {Miski-Oglu},\ and\ \citenamefont {Richter}}]{Bittner2013b}%
  \BibitemOpen
  \bibfield  {author} {\bibinfo {author} {\bibfnamefont {S.}~\bibnamefont
  {Bittner}}, \bibinfo {author} {\bibfnamefont {E.}~\bibnamefont {Bogomolny}},
  \bibinfo {author} {\bibfnamefont {B.}~\bibnamefont {Dietz}}, \bibinfo
  {author} {\bibfnamefont {M.}~\bibnamefont {Miski-Oglu}}, \ and\ \bibinfo
  {author} {\bibfnamefont {A.}~\bibnamefont {Richter}},\ }\href {\doibase
  10.1103/PhysRevE.88.062906} {\bibfield  {journal} {\bibinfo  {journal} {Phys.
  Rev. E}\ }\textbf {\bibinfo {volume} {88}},\ \bibinfo {pages} {062906}
  (\bibinfo {year} {2013})}\BibitemShut {NoStop}%
\bibitem [{\citenamefont {Bittner}\ \emph {et~al.}(2014)\citenamefont
  {Bittner}, \citenamefont {Bogomolny}, \citenamefont {Dietz}, \citenamefont
  {Miski-Oglu},\ and\ \citenamefont {Richter}}]{Bittner2014a}%
  \BibitemOpen
  \bibfield  {author} {\bibinfo {author} {\bibfnamefont {S.}~\bibnamefont
  {Bittner}}, \bibinfo {author} {\bibfnamefont {E.}~\bibnamefont {Bogomolny}},
  \bibinfo {author} {\bibfnamefont {B.}~\bibnamefont {Dietz}}, \bibinfo
  {author} {\bibfnamefont {M.}~\bibnamefont {Miski-Oglu}}, \ and\ \bibinfo
  {author} {\bibfnamefont {A.}~\bibnamefont {Richter}},\ }\href {\doibase
  10.1103/PhysRevE.90.052909} {\bibfield  {journal} {\bibinfo  {journal} {Phys.
  Rev. E}\ }\textbf {\bibinfo {volume} {90}},\ \bibinfo {pages} {052909}
  (\bibinfo {year} {2014})}\BibitemShut {NoStop}%
\bibitem [{\citenamefont {Lozenko}\ \emph {et~al.}(2012)\citenamefont
  {Lozenko}, \citenamefont {Djellali}, \citenamefont {Gozhyk}, \citenamefont
  {Delezoide}, \citenamefont {Lau\-tru}, \citenamefont {Ulysse}, \citenamefont
  {Zyss},\ and\ \citenamefont {Lebental}}]{Lozenko2012}%
  \BibitemOpen
  \bibfield  {author} {\bibinfo {author} {\bibfnamefont {S.}~\bibnamefont
  {Lozenko}}, \bibinfo {author} {\bibfnamefont {N.}~\bibnamefont {Djellali}},
  \bibinfo {author} {\bibfnamefont {I.}~\bibnamefont {Gozhyk}}, \bibinfo
  {author} {\bibfnamefont {C.}~\bibnamefont {Delezoide}}, \bibinfo {author}
  {\bibfnamefont {J.}~\bibnamefont {Lau\-tru}}, \bibinfo {author}
  {\bibfnamefont {C.}~\bibnamefont {Ulysse}}, \bibinfo {author} {\bibfnamefont
  {J.}~\bibnamefont {Zyss}}, \ and\ \bibinfo {author} {\bibfnamefont
  {M.}~\bibnamefont {Lebental}},\ }\href {\doibase 10.1063/1.4720474}
  {\bibfield  {journal} {\bibinfo  {journal} {J. Appl. Phys.}\ }\textbf
  {\bibinfo {volume} {111}},\ \bibinfo {pages} {103116} (\bibinfo {year}
  {2012})}\BibitemShut {NoStop}%
\bibitem [{\citenamefont {Lebental}\ \emph {et~al.}(2007)\citenamefont
  {Lebental}, \citenamefont {Djellali}, \citenamefont {Arnaud}, \citenamefont
  {Lauret}, \citenamefont {Zyss}, \citenamefont {Dubertrand}, \citenamefont
  {Schmit},\ and\ \citenamefont {Bogomolny}}]{Lebental2007}%
  \BibitemOpen
  \bibfield  {author} {\bibinfo {author} {\bibfnamefont {M.}~\bibnamefont
  {Lebental}}, \bibinfo {author} {\bibfnamefont {N.}~\bibnamefont {Djellali}},
  \bibinfo {author} {\bibfnamefont {C.}~\bibnamefont {Arnaud}}, \bibinfo
  {author} {\bibfnamefont {J.-S.}\ \bibnamefont {Lauret}}, \bibinfo {author}
  {\bibfnamefont {J.}~\bibnamefont {Zyss}}, \bibinfo {author} {\bibfnamefont
  {R.}~\bibnamefont {Dubertrand}}, \bibinfo {author} {\bibfnamefont
  {C.}~\bibnamefont {Schmit}}, \ and\ \bibinfo {author} {\bibfnamefont
  {E.}~\bibnamefont {Bogomolny}},\ }\href {\doibase 10.1103/PhysRevA.76.023830}
  {\bibfield  {journal} {\bibinfo  {journal} {Phys. Rev. A}\ }\textbf {\bibinfo
  {volume} {76}},\ \bibinfo {pages} {023830} (\bibinfo {year}
  {2007})}\BibitemShut {NoStop}%
\bibitem [{\citenamefont {Bogomolny}\ \emph {et~al.}(2011)\citenamefont
  {Bogomolny}, \citenamefont {Djellali}, \citenamefont {Dubertrand},
  \citenamefont {Gozhyk}, \citenamefont {Lebental}, \citenamefont {Schmit},
  \citenamefont {Ulysse},\ and\ \citenamefont {Zyss}}]{Bogomolny2011}%
  \BibitemOpen
  \bibfield  {author} {\bibinfo {author} {\bibfnamefont {E.}~\bibnamefont
  {Bogomolny}}, \bibinfo {author} {\bibfnamefont {N.}~\bibnamefont {Djellali}},
  \bibinfo {author} {\bibfnamefont {R.}~\bibnamefont {Dubertrand}}, \bibinfo
  {author} {\bibfnamefont {I.}~\bibnamefont {Gozhyk}}, \bibinfo {author}
  {\bibfnamefont {M.}~\bibnamefont {Lebental}}, \bibinfo {author}
  {\bibfnamefont {C.}~\bibnamefont {Schmit}}, \bibinfo {author} {\bibfnamefont
  {C.}~\bibnamefont {Ulysse}}, \ and\ \bibinfo {author} {\bibfnamefont
  {J.}~\bibnamefont {Zyss}},\ }\href {\doibase 10.1103/PhysRevE.83.036208}
  {\bibfield  {journal} {\bibinfo  {journal} {Phys. Rev. E}\ }\textbf {\bibinfo
  {volume} {83}},\ \bibinfo {pages} {036208} (\bibinfo {year}
  {2011})}\BibitemShut {NoStop}%
\bibitem [{\citenamefont {Gozhyk}\ \emph {et~al.}(2012)\citenamefont {Gozhyk},
  \citenamefont {Clavier}, \citenamefont {M\'eallet-Renault}, \citenamefont
  {Dvor\-ko}, \citenamefont {Pansu}, \citenamefont {Audibert}, \citenamefont
  {Brosseau}, \citenamefont {Lafargue}, \citenamefont {Tsvirkun}, \citenamefont
  {Lo\-zenko}, \citenamefont {Forget}, \citenamefont {Ch\'enais}, \citenamefont
  {Ulysse}, \citenamefont {Zyss},\ and\ \citenamefont {Lebental}}]{Gozhyk2012}%
  \BibitemOpen
  \bibfield  {author} {\bibinfo {author} {\bibfnamefont {I.}~\bibnamefont
  {Gozhyk}}, \bibinfo {author} {\bibfnamefont {G.}~\bibnamefont {Clavier}},
  \bibinfo {author} {\bibfnamefont {R.}~\bibnamefont {M\'eallet-Renault}},
  \bibinfo {author} {\bibfnamefont {M.}~\bibnamefont {Dvor\-ko}}, \bibinfo
  {author} {\bibfnamefont {R.}~\bibnamefont {Pansu}}, \bibinfo {author}
  {\bibfnamefont {J.-F.}\ \bibnamefont {Audibert}}, \bibinfo {author}
  {\bibfnamefont {A.}~\bibnamefont {Brosseau}}, \bibinfo {author}
  {\bibfnamefont {C.}~\bibnamefont {Lafargue}}, \bibinfo {author}
  {\bibfnamefont {V.}~\bibnamefont {Tsvirkun}}, \bibinfo {author}
  {\bibfnamefont {S.}~\bibnamefont {Lo\-zenko}}, \bibinfo {author}
  {\bibfnamefont {S.}~\bibnamefont {Forget}}, \bibinfo {author} {\bibfnamefont
  {S.}~\bibnamefont {Ch\'enais}}, \bibinfo {author} {\bibfnamefont
  {C.}~\bibnamefont {Ulysse}}, \bibinfo {author} {\bibfnamefont
  {J.}~\bibnamefont {Zyss}}, \ and\ \bibinfo {author} {\bibfnamefont
  {M.}~\bibnamefont {Lebental}},\ }\href {\doibase 10.1103/PhysRevA.86.043817}
  {\bibfield  {journal} {\bibinfo  {journal} {Phys. Rev. A}\ }\textbf {\bibinfo
  {volume} {86}},\ \bibinfo {pages} {043817} (\bibinfo {year}
  {2012})}\BibitemShut {NoStop}%
\bibitem [{\citenamefont {Ben-Messaoud}\ and\ \citenamefont
  {Zyss}(2005)}]{Ben-Messaoud2005}%
  \BibitemOpen
  \bibfield  {author} {\bibinfo {author} {\bibfnamefont {T.}~\bibnamefont
  {Ben-Messaoud}}\ and\ \bibinfo {author} {\bibfnamefont {J.}~\bibnamefont
  {Zyss}},\ }\href {\doibase DOI:10.1063/1.1949708} {\bibfield  {journal}
  {\bibinfo  {journal} {Appl. Phys. Lett.}\ }\textbf {\bibinfo {volume} {86}},\
  \bibinfo {pages} {241110} (\bibinfo {year} {2005})}\BibitemShut {NoStop}%
\bibitem [{\citenamefont {Bachelard}\ \emph {et~al.}(2012)\citenamefont
  {Bachelard}, \citenamefont {Andreasen}, \citenamefont {Gigan},\ and\
  \citenamefont {Sebbah}}]{Bachelard2012}%
  \BibitemOpen
  \bibfield  {author} {\bibinfo {author} {\bibfnamefont {N.}~\bibnamefont
  {Bachelard}}, \bibinfo {author} {\bibfnamefont {J.}~\bibnamefont
  {Andreasen}}, \bibinfo {author} {\bibfnamefont {S.}~\bibnamefont {Gigan}}, \
  and\ \bibinfo {author} {\bibfnamefont {P.}~\bibnamefont {Sebbah}},\ }\href
  {\doibase 10.1103/PhysRevLett.109.033903} {\bibfield  {journal} {\bibinfo
  {journal} {Phys. Rev. Lett.}\ }\textbf {\bibinfo {volume} {109}},\ \bibinfo
  {pages} {033903} (\bibinfo {year} {2012})}\BibitemShut {NoStop}%
\bibitem [{\citenamefont {Liew}\ \emph {et~al.}(2014)\citenamefont {Liew},
  \citenamefont {Redding}, \citenamefont {Ge}, \citenamefont {Solomon},\ and\
  \citenamefont {Cao}}]{Liew2014}%
  \BibitemOpen
  \bibfield  {author} {\bibinfo {author} {\bibfnamefont {S.~F.}\ \bibnamefont
  {Liew}}, \bibinfo {author} {\bibfnamefont {B.}~\bibnamefont {Redding}},
  \bibinfo {author} {\bibfnamefont {L.}~\bibnamefont {Ge}}, \bibinfo {author}
  {\bibfnamefont {G.~S.}\ \bibnamefont {Solomon}}, \ and\ \bibinfo {author}
  {\bibfnamefont {H.}~\bibnamefont {Cao}},\ }\href {\doibase 10.1063/1.4883637}
  {\bibfield  {journal} {\bibinfo  {journal} {Appl. Phys. Lett.}\ }\textbf
  {\bibinfo {volume} {104}},\ \bibinfo {pages} {231108} (\bibinfo {year}
  {2014})}\BibitemShut {NoStop}%
\bibitem [{\citenamefont {Wiersig}(2003{\natexlab{b}})}]{Wiersig2003a}%
  \BibitemOpen
  \bibfield  {author} {\bibinfo {author} {\bibfnamefont {J.}~\bibnamefont
  {Wiersig}},\ }\href {\doibase 10.1088/1464-4258/5/1/308} {\bibfield
  {journal} {\bibinfo  {journal} {J. Opt. A}\ }\textbf {\bibinfo {volume}
  {5}},\ \bibinfo {pages} {53} (\bibinfo {year}
  {2003}{\natexlab{b}})}\BibitemShut {NoStop}%
\bibitem [{\citenamefont {Peres}(1984)}]{Peres1984}%
  \BibitemOpen
  \bibfield  {author} {\bibinfo {author} {\bibfnamefont {A.}~\bibnamefont
  {Peres}},\ }\href {\doibase 10.1103/PhysRevLett.53.1711} {\bibfield
  {journal} {\bibinfo  {journal} {Phys. Rev. Lett.}\ }\textbf {\bibinfo
  {volume} {53}},\ \bibinfo {pages} {1711} (\bibinfo {year}
  {1984})}\BibitemShut {NoStop}%
\end{thebibliography}
%

% Begin Supplementary Material
% --------------------------------------------
\renewcommand{\theequation}{S\arabic{equation}}
\renewcommand{\thefigure}{S\arabic{figure}}
\setcounter{figure}{0}
\setcounter{equation}{0}

\section{Supplementary Material: Calculation of the far-field distributions}
We use Green's identity, 
\begin{equation} \Psi(\vec{r}) = \oint_{\partial S} \, d|\vec{r}{\,'}| \left\{ \, G(\vec{r}, \vec{r}{\,'}) \pbyp{\Psi}{n}(\vec{r}{\,'}) \right. \left. - \pbyp{G}{n}(\vec{r}, \vec{r}{\,'}) \Psi(\vec{r}{\,'}) \right\} \, , \end{equation}
to infer the far-field distribution (cf.\ Ref.~\cite{Yang2007c}), where $\pbyp{}{n}$ is the derivative with respect to the surface normal of the domain $S$ and
\begin{equation} G(\vec{r}, \vec{r}{\,'}) = H_0^{(1)}(k |\vec{r} - \vec{r}{\,'}|) / (4 i) \end{equation}
is the Green's function in two dimensions with $k = (k_x^2 + k_y^2)^{1/2} / n$ being the wave number. We choose the boundary of the resonator as the integration path $\partial S$ and assume continuity of $\Psi$ and $\mu \pbyp{\Psi}{n}$ at the resonator boundaries where $\mu = 1$ ($\mu = 1 / n^2$) for TM (TE) polarization \cite{Lebental2007}. Then the wave function outside of the resonator, $\Psi_\mathrm{out}$, is related to that inside, $\Psi_\mathrm{in}$, by 
\begin{equation} \Psi_\mathrm{out}(\vec{r}) = \oint_{\partial S} \, d|\vec{r}{\,'}| \left\{ \, \mu G \pbyp{\Psi_\mathrm{in}}{n} - \pbyp{G}{n} \Psi_\mathrm{in} \right\} \, . \end{equation}
We start with a single plane wave $\Phi_\mathrm{in}(\vec{r}) = \Phi_0 \exp\{i (k_x x + k_y y) \}$ inside the resonator and calculate the corresponding outside wave function $\Phi_\mathrm{out}$ in the far field. The integral can be simplified in this case where $x = r \cos(\varphi)$ and $y = r \sin(\varphi)$ with $r \rightarrow \infty$ by using $H_0^{(1)}(z) \approx \sqrt{2 / (\pi z)} e^{i (z - \pi / 4)}$ for $|z| = k |\vec{r} - \vec{r}{\,'}| \rightarrow \infty$. Furthermore, in this limit $|\vec{r} - \vec{r}{\,'}| \simeq r - x' \cos(\varphi) - y' \sin(\varphi)$. The integration yields
\begin{widetext}
\begin{equation} \Phi_\mathrm{out}(r, \varphi) = -i \Phi_0 \sqrt{\frac{2}{\pi k r}} e^{i(k r - \pi / 4)} \sin\left[(k_x - k \cos{\varphi}) \frac{a}{2}\right] \sin\left[(k_y - k \sin{\varphi}) \frac{a}{2}\right] \left[ \frac{\mu k_x + k \cos{\varphi}}{k_y - k \sin{\varphi}} + \frac{\mu k_y + k \sin{\varphi}}{k_x - k \cos{\varphi}} \right] \, . \end{equation}
It is instructive to replace $k_x = n k \cos{\alphainc}$ and $k_y = n k \sin{\alphainc}$ and look at the maxima of the two terms in the large square brackets. The first term stems from the integration along the side walls parallel to the $y$ axis and is maximal when its denominator vanishes, i.e., for $n \sin{\alphainc} = \sin{\varphi}$. This is simply Snell's law for a ray refracted at a side wall parallel to the $y$ axis, and analogously the maximum of the second term corresponds to a ray refracted at a side wall parallel to the $x$ axis. % paragraph

The results for the full wave functions of the dielectric square are obtained by adding up the $8$ plane waves with their correct momentum vectors and relative amplitudes. For the symmetry class $A_2$, this yields
\begin{equation} \label{eq:ffModel} \Psi_{A_2}(r, \varphi) = -i \frac{\Psi_0}{2 \sqrt{2 \pi k r}} e^{i(k r - \pi / 4)} g_{A_2}(\varphi) \end{equation}
with
\begin{equation} \scriptscriptstyle \label{eq:ffA2} \begin{array}{rcl} g_{A_2}(\varphi) & = & 2 \{ \sinc[(k_y - k \sin{\varphi}) \frac{a}{2}] - \sinc[(-k_y - k \sin{\varphi}) \frac{a}{2}] \} \\
& & \times [\mu k_x \frac{a}{2} \cos(k_x \frac{a}{2}) \sin(k \frac{a}{2} \cos{\varphi}) - k \frac{a}{2} \cos{\varphi} \cos(k \frac{a}{2} \cos{\varphi}) \sin(k_x \frac{a}{2})] \\ \\
& & +2 \{ \sinc[(k_x - k \cos{\varphi}) \frac{a}{2}] - \sinc[(-k_x - k \cos{\varphi}) \frac{a}{2}] \} \\
& & \times [\mu k_y \frac{a}{2} \cos(k_y \frac{a}{2}) \sin(k \frac{a}{2} \sin{\varphi}) - k \frac{a}{2} \sin{\varphi} \cos(k \frac{a}{2} \sin{\varphi}) \sin(k_y \frac{a}{2})] \\ \\
& & +2 \{ -\sinc[(k_x - k \sin{\varphi}) \frac{a}{2}] + \sinc[(-k_x - k \sin{\varphi}) \frac{a}{2}] \} \\
& & \times [\mu k_y \frac{a}{2} \cos(k_y \frac{a}{2}) \sin(k \frac{a}{2} \cos{\varphi}) - k \frac{a}{2} \cos{\varphi} \cos(k \frac{a}{2} \cos{\varphi}) \sin(k_y \frac{a}{2})] \\ \\
& & +2 \{ -\sinc[(k_y - k \cos{\varphi}) \frac{a}{2}] + \sinc[(-k_y - k \cos{\varphi}) \frac{a}{2}] \} \\
& & \times [\mu k_x \frac{a}{2} \cos(k_x \frac{a}{2}) \sin(k \frac{a}{2} \sin{\varphi}) - k \frac{a}{2} \sin{\varphi} \cos(k \frac{a}{2} \sin{\varphi}) \sin(k_x \frac{a}{2})] \, . \\ \\
\end{array} \end{equation}
\end{widetext}

\section{Supplementary Material: Comparison with numerical calculations}

% Figure S1: Numerical spectrum re-vs-alpha
\begin{figure}[tb]
\begin{center}
\includegraphics[width = 8.4 cm]{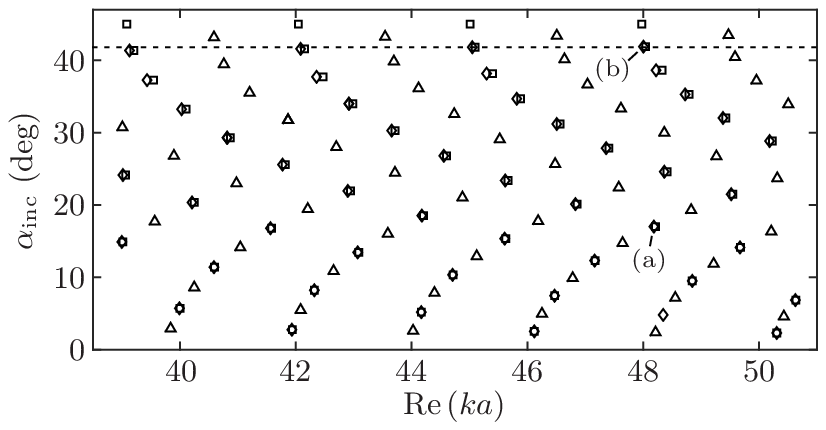}
\end{center}
\caption{Numerically calculated spectrum of a square microlaser with $n = 1.5$. The angle of incidence characterizing the modes is plotted with respect to the real part of $ka$. The different symbols correspond to the symmetry classes $(--)$ (diamonds), $(++)$ (squares), and the degenerate classes $(-+)$ and $(+-)$ (triangles). The dashed horizontal line indicates the critical angle. The far-field intensity distributions of the $(--)$ modes labeled (a) and (b) are presented in \reffig{fig:FFnum}.}
\label{fig:specNum}
\end{figure}

To validate this semiclassical model, numerical simulations were performed using the boundary element method \cite{Wiersig2003a, Lebental2007}. The complete spectrum of TM modes as well as the near- and far-field distributions were calculated in the range of $\Re{ka} \approx 38$--$51$ for $n = 1.5$. Figure \ref{fig:specNum} shows the numerically calculated spectrum. For each resonance, we inferred the quantum numbers $m_x$ and $m_y$, and then the corresponding $\alphainc$, by comparing the near-field distributions to the model predictions. The very regular structure of the spectrum and the possibility to quantize it through $m_x$ and $m_y$ confirm that the dielectric square resonator is nearly an integrable system \cite{Peres1984, Bittner2014a}. 

% Figure S2: Two examples of numerical far-field distributions
\begin{figure}[tb]
\begin{center}
\subfigure[]{
	\includegraphics[width = 5.5 cm]{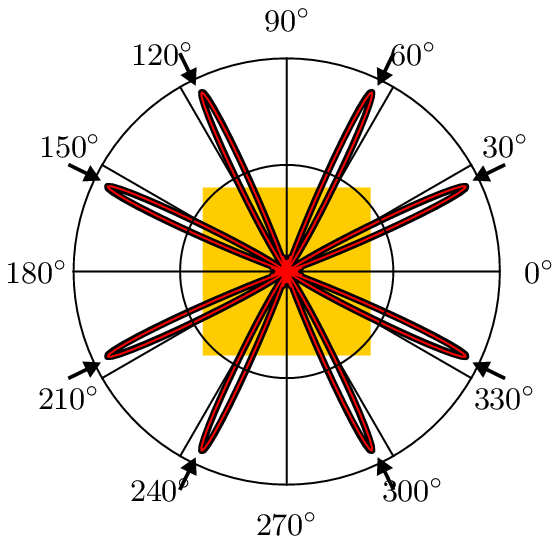}
	\label{sfig:FFnumA}
}
\subfigure[]{
	\includegraphics[width = 5.5 cm]{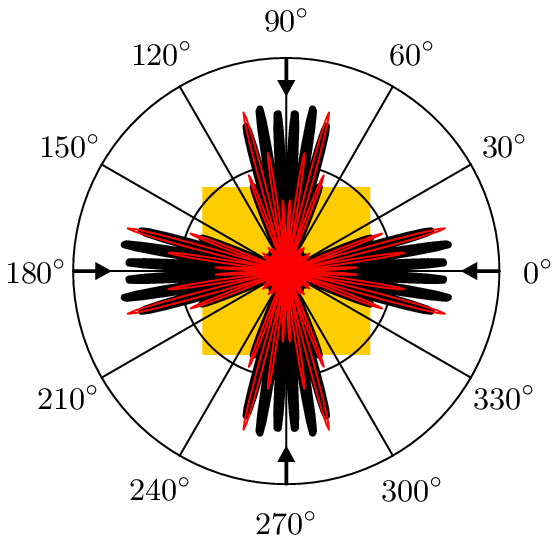}
	\label{sfig:FFnumB}
}
\end{center}
\caption{(Color online) Numerical (black) and model (red) calculations of the far-field intensity distributions of resonances (a) TM$(6, 22, --)$ with $\alphainc = 17.0^\circ$ and (b) TM$(15, 17, --)$ with $\alphainc = 41.9^\circ$. The arrows indicate the directions of the corresponding refracted rays.}
\label{fig:FFnum}
\end{figure}

Figure \ref{fig:FFnum} presents typical far-field distributions, one with $\alphainc$ well below $\alphacrit$ [panel (a)], and the other with $\alphainc$ just above $\alphacrit$ [panel (b)]. In Fig.~S\ref{sfig:FFnumA}, the mode exhibits at total of $8$ narrow refractive emission lobes. Their directions agree very well with those expected from ray optics indicated by the arrows. The analytically calculated far-field distribution (red line) agrees almost perfectly with the numerical calculation (black line). The $8$ nonrefractive emission lobes cannot be recognized since their amplitudes are negligible compared to those of the refractive lobes. It should be noted that the lobes of this example are much broader than those of the corresponding example in Fig.~6(c) of the main text for larger $ka$ since the ratio between the effective slit size $a_\mathrm{eff}$ and the wavelength is much smaller here. In Fig.~S\ref{sfig:FFnumB}, the directions of the various lobes are very well reproduced by the model although it sometimes fails to predict their amplitude correctly. In general, the ray-based model is less accurate for modes with $\alphainc \gtrsim \alphacrit$ since the Fresnel reflection coefficients for an infinite interface that are used in the quantization condition are no longer a good approximation close to and above the critical angle. This is evidenced by the prediction of zero losses for such modes by the model, whereas all resonances naturally have a finite $\Im{ka} < 0$ in reality. The nonrefractive lobes are no longer narrow but broad and jagged. The model has only been tested for moderately sized cavities [$\Re{ka} \leq 60$], and not for $\Re{ka} = 800$--$2000$ like those investigated experimentally, because simulations in this size region are extremely cumbersome. However, thanks to the semiclassical nature of the model, it seems reasonable that it works well for very large cavities.

\section{Supplementary Material: Envelope of the far-field distributions}
The fast oscillations of the far-field intensity distribution are not visible with an angular resolution of $1^\circ$, and the measured far-field patterns are hence compared with the envelope of the model far-field distributions. The terms in the square brackets in \refeq{eq:ffA2}, for example the first term of the equation,
\begin{equation} \begin{array}{rcl} h_{A_2, 1}(\varphi) & = & \mu k_x \frac{a}{2} \cos(k_x a / 2) \sin(k a \cos{\varphi} / 2) \\ & & - k \frac{a}{2} \cos{\varphi} \cos(k a \cos{\varphi} / 2) \sin(k_x a / 2) \, , \end{array} \end{equation}
feature these very fast oscillations due to the $\sin$ and $\cos$ functions with the argument $\phi = k a \cos(\varphi) / 2$. The envelope of $h_{A_2, 1}$ is calculated by considering $\phi$ as an independent variable and replacing it by the value 
\begin{equation} \phi_0(\varphi) = - \mathrm{arctan} \left\{ \frac{\mu k_x}{k \cos{\varphi} \tan(k_x a / 2)} \right\} \end{equation}
for which $h_{A_2, 1}$ becomes extremal, yielding
\begin{equation} \begin{array}{rcl} h_{A_2, 1}^\mathrm{(env)}(\varphi) & = & \mu k_x \frac{a}{2} \cos(k_x a / 2) \sin(\phi_0) - \\ & & k \frac{a}{2} \cos{\varphi} \cos(\phi_0) \sin(k_x a / 2) \, . \\ \\ \end{array} \end{equation}
Analogous formulas are obtained for the other terms. The function that is fitted to the experimental data can be further simplified since for the parameter range of interest only one of the $\sinc$ terms contributes significantly to each emission lobe. So for example the lobe around $\varphi = 90^\circ$ is fitted by the formula
\begin{equation} \label{eq:Ifit} I_\mathrm{fit}(\varphi) = A \left| 2 \, \sinc \left[(k_y - k \sin{\varphi}) \frac{a}{2} \right] h_{A_2, 1}^\mathrm{(env)}(\varphi) \right|^2 \end{equation}
where the only two fit parameters are the amplitude $A$ and the angle of incidence $\alphainc$ which is related to the momentum vector components via $k_x = n k \cos{\alphainc}$ and $k_y = n k \sin{\alphainc}$. % paragraph

Each lobe of the measured far-field intensity distribution was fitted separately in the range $[q \pi / 2 - \pi / 6, q \pi / 2 + \pi / 6]$ where $q = 1, \, 2, \, 3, \, 4$ by \refeq{eq:Ifit} or a corresponding expression. TE polarization was used. The envelopes of the far-field distributions are almost identical for the different symmetry classes and hence fitting the formula for another symmetry class to the experimental data yields nearly the same parameters and an equally good agreement. The effective refractive index of $n = 1.50$ was assumed to be a constant since the shape of the envelope depends very little on it, and the fits are insofar independent of $n$ as the fit yields practically the same value for $\alphainc^\mathrm{(fit)} - \alphacrit$ for a wide range of values of $n$.

\end{document}